\documentclass[conference]{IEEEtran}
\IEEEoverridecommandlockouts
\usepackage{cite}
\usepackage{amsmath,amssymb,amsfonts}
\usepackage{algorithmic}
\usepackage{graphicx}
\usepackage{textcomp}
\usepackage{xcolor}

\makeatletter 
\newcommand{\linebreakand}{%
  \end{@IEEEauthorhalign}
  \hfill\mbox{}\par
  \mbox{}\hfill\begin{@IEEEauthorhalign}
}
\makeatother 

\def\BibTeX{{\rm B\kern-.05em{\sc i\kern-.025em b}\kern-.08em
    T\kern-.1667em\lower.7ex\hbox{E}\kern-.125emX}}
\begin{document}

\title{An Open-Ended Approach to Understanding Local, Emergent Conservation Laws in Biological Evolution\\
\thanks{This work was funded by Cross Compass, LTD.}
}

\author{
\IEEEauthorblockN{Alyssa M Adams}
\IEEEauthorblockA{\textit{Cross Labs, Cross Compass}\\
Kyoto, Japan\\
alyssa.adams@cross-compass.com}
\and
\IEEEauthorblockN{Eliott Jacopin}
\IEEEauthorblockA{\textit{Center for Biosystems Dynamics Research}
\textit{RIKEN}\\
Kobe, Japan \\
http://orcid.org/0000-0002-4568-283X}
\and
\IEEEauthorblockN{Praful Gagrani}
\IEEEauthorblockA{\textit{Wisconsin Institute for Discovery}\\
\textit{University of Wisconsin-Madison}\\
Madison, WI, USA}
\and

\linebreakand

\IEEEauthorblockN{Olaf Witkowski}
\IEEEauthorblockA{\textit{Cross Labs, Cross Compass}\\
Kyoto, Japan\\
olaf@cross-compass.com}
}

\maketitle

\begin{abstract}

While fields like Artificial Life have made huge strides in quantifying the mechanisms that distinguish living systems from non-living ones, particular mechanisms remain difficult to reproduce \textit{in silico}. In particular, biology seems to endlessly produce new, innovative states and new, innovative rules. Known as open-endedness, we've been successful in finding mechanisms that generate new states, but have been less successful in finding mechanisms that generate new \textit{rules}. In this discussion paper, we weigh whether or not analyzing the effects of internal and external system constraints on a system's dynamics would be a fruitful avenue to understanding open-endedness in future studies. We discuss the conceptual connection between physical constraints and the ways that the system can physically reach possible states while those constraints are present. It seems that the physical constraints that define biological objects (and dynamics) are maintained by dynamics that occur from \textit{within the system}. This is in opposition to current modeling approaches where system constraints are maintained by external forces. We suggest that constraints can be characterized as variables whose values are either completely conserved, quasi-conserved, or conditionally conserved. Regardless of whether or not a constrained variable is a part of the biological object or present in the object's environment, we discuss how the accessible system states under that constraint can lead to local, emergent conservation laws (rules). We provide several examples of constraints based on real, well-studied systems and highlight the differences in mechanisms that constrain the variables being held constant. Finally, we discuss the possible benefits of formally understanding how system constraints that emerge \textit{from within} a system lead to system dynamics that can be characterized as new, emergent rules-- particularly for artificial intelligence, hybrid life, embodiment, astrobiology, and more. Understanding how new, local rules might emerge \textit{from within the system} is crucial for understanding how open-ended systems continually discover new update rules, in addition to continually discovering new states.

\end{abstract}


\section{Introduction}














Over the last few decades, researchers have made much progress in identifying aspects of living systems that set them apart from non-living ones \cite{cite:Taylor_2021, cite:Bartlett_Wong, cite:Wong_Bartlett_Chen_Tierney, cite:Wong_etal, cite:Sharma, cite:Kim, cite:Kempes, cite:Dupre, cite:Witzany, cite:Krakauer, cite:Jeancolas, cite:Pattee, cite:Packard, cite:Flack, cite:Hernandez-Orozco, cite:Zenil}, even if there is still much more to be learned \cite{cite:stanley, cite:Song, cite:Packard19, cite:Bedau, cite:stanley17, cite:Taylor_2016, cite:Soros}. One popular approach, particularly in Artificial Life (ALife), is to replicate some aspects of living systems in a computer simulation. By doing so, it becomes possible to tinker with the computer code and hardware to explore different system behaviors. Based on these results, researchers can make connections between their experiment's design, and real, biological processes.

Many mechanisms seem to drive living systems. But they have been difficult to quantify, measure, and understand well enough that they can be replicated in a computer program. One aspect of life, in particular, continues to be difficult to understand-- its ability to continually find and select new, previously unexplored states and dynamics \textit{without} external intervention \cite{cite:Taylor_2016, cite:Packard19, cite:Song}. Known as open-endedness, this ``behavior'' continuously produces new, innovative states and evolves under new, innovative dynamics without repeating itself exactly over time \cite{cite:Taylor_2016, cite:Packard19, cite:Song, cite:banzhaf}. Theoretical approaches to exactly define open-endedness have faced an inability to arrive at a field-wide consensus \cite{cite:Taylor_2016, cite:Packard19, cite:Song}. So while it is understood in terms of broad, qualitative definitions, an exact, mathematical definition that applies to any dynamic system remains to be decided.

Computationally, engineering and quantifying open-ended dynamics in a closed system has been particularly challenging \cite{cite:Taylor_2016, cite:Hintze}. Because there isn't a field-wide standard mathematical definition to detect or quantify ``how much open-endedness'' or ``what type of open-endedness'' is in a system \cite{cite:Song, cite:Hintze}, it is unclear how to interpret novel models that produce interesting dynamics (\cite{cite:Wang, cite:Wang20, cite:Dolson, cite:Horibe}, to name a just a few). But it is also unclear how open-ended systems continue to produce new rules as a result of dynamics and interactions \textit{from within} a system, as opposed to constant, external interventions being \textit{made on} the system \cite{cite:Taylor_2021}.

For bounded, deterministic computational approaches, implementing systems that continually find new, innovative rules to evolve under is a challenge. As has been suggested previously by others \cite{cite:Ackley, cite:Ackley14}, we suggest that a big part of this challenge is due to the way we approach modeling. Current approaches involve manually writing a set of rules (a program) with tunable variables (parameters), running the program on a set of hardware, and characterizing the resulting dynamics of the system. The program, parameters, and hardware are either all manually designed, or can be set to change as some function of the system's state, dynamics, and/or external parameters \cite{cite:PyMC, cite:Gen, cite:Adams}. 

There already exist a few explicit methods of changing the set of rules within a system over time. In probabilistic computing, a data-prediction algorithm can be edited and changed by a secondary ``meta-program.'' But even if the lower-level program is continually changing, the meta-program that controls doesn't change directly as a result of ``within-the-system'' dynamics. The resulting dynamics of the changing program are still a direct result of an external force acting externally on the system.

So then, how do rules in systems that evolve open-endedly change as a result of dynamics and interactions \textit{from within} the system \cite{cite:Taylor_2021}? Moreover, how can we understand biological rules like autopoesis, homeostasis, metabolism, replication, etc. in terms of local, emergent conservation laws (or, from another perspective, dynamics that \textit{could} be described as new, emergent updates rules) acting from within the system, even if not defined explicitly from global update laws (such as the laws of physics)? Is it possible to do so starting from global laws like physics and chemistry? How do internal and external system constraints change the system dynamics in a way that they appear to evolve under new, emergent rules? 

Understanding how new, local ``rules'' might emerge \textit{from within the system} is crucial for understanding how open-ended systems continually discover new update rules, in addition to continually discovering new states. \textbf{The purpose of this paper is to weigh whether or not analyzing the effects of internal and external system constraints on a system's dynamics would be a fruitful avenue to understanding open-endedness in future studies.} We start by outlining some of the current challenges in ALife like open-endedness that are impacted by these problems. We review current ALife approaches on how open-endedly changing rules might build on each other to create the kind of ``ratcheting effect'' we see in in the evolution of various systems (biological, cultural, etc.) \cite{cite:Liard, cite:Tennie}. Next, we briefly review the characterization of the most important rules that distinguish living systems (``lyfe'') from abiotic ones (``non-lyfe'') \cite{cite:Bartlett_Wong, cite:Wong_Bartlett_Chen_Tierney}. Finally, we discuss future directions and possible implications of a ``constraint-centric'' perspective of open-endedness for ALife, artificial intelligence (AI), hybrid life, major transitions in biological evolution, and astrobiology \cite{cite:Taylor_2016, cite:Packard19, cite:Song, cite:Baltieri}. 

\section{Challenges in Artificial Life}

Understanding how biology evolves open-endedly, and what the mechanisms for open-endedness are more broadly, remains an open challenge for ALife \cite{cite:Smith}. In general, open-endedness can be classified into different types and characterized in different kinds of systems \cite{cite:Taylor_2019, cite:Packard, cite:Packard19, cite:Bedau, cite:Bedau97, cite:Bedauetal97}. It is generally agreed upon that any system that continually explores new states \textit{and also} continually explores new ways to find those states is open-ended to some degree \cite{cite:Taylor_2016, cite:Packard19, cite:Song, cite:banzhaf}. But that degree can vary wildly from system to system.

These challenges are at the intersection of ALife machine learning \cite{cite:Bartlett_Louapre, cite:stanley17, cite:stanley}, astrobiology \cite{cite:banzhaf}, economics \cite{cite:stanley17, cite:stanley}, mathematics \cite{cite:Orozco, cite:Lavin}, and complex systems \cite{cite:Mitchell}. The implications for fully understanding and being able to engineer the necessary and sufficient mechanisms that produce open-ended behavior in a system are profound.

\subsection{Emergence}

The concept of emergence (such as the emergence of new system behavior, the emergence of new kinds of objects, etc) is often at the center of many problems in ALife \cite{cite:Bedau97}. The standard approach involves showing how new rules (new, ``effective laws of physics'') emerge as a result of lower-level, finer-grained dynamics of ``particles.''

Some examples of particles from physics include defects in a crystal lattice and vortices in a fluid. The equations that describe their behavior, such as potentials and motion, treat the objects as if they are fundamental particles. But those equations are mediated by the collective action of many, many sub-particles at a lower level. In these cases, ``new physics'' emerges from the underlying dynamics, each with its own apparent rules and conservation laws that are not explicitly part of the symmetries of the underlying physics.

In a crystal lattice, the 'defects' are not conserved because they're invariants of the Hamiltonian of the universe. They're conserved because removing them would involve rearranging the entire crystal, and the crystal as a whole is much bigger than the defect; it's an event that happens with (asymptotic with system size) probability zero. Therefore, the defect becomes a new emergent conservation law of the coarse-grained system even if the variable it tracks (defined on the coarse-grained macro-scale) doesn't even exist at the micro-scale.

\subsection{The continual creation of new, local rules}

Overall, we have a good understanding of static mechanisms that continually produce new states. For example, counting to infinity is a static mechanism (always add 1 to the previous number) that always guarantees a new, never-before-seen state. But understanding internal processes that continually generate new, never-before-seen rules is much less understood \cite{cite:Taylor_2019, cite:Packard, cite:Packard19, cite:Bedau, cite:Bedau97, cite:Bedauetal97, cite:Adams}. 

Beyond the importance of recreating open-ended systems computationally, knowing how new rules emerge (and persist) is crucial for understanding major evolutionary transitions, including the transition from a non-living state to a living one. Each major transition is marked by a new kind of dynamical rule that persists across several, future biological objects \cite{cite:Taylor_2019, cite:Packard, cite:Packard19, cite:Bartlett_Louapre, cite:Montevil_Mossio}. 

Computationally we already implement systems that change a system's program (set of rules) as a result of a secondary meta-rule. In machine learning, training a model involves iterating over different individual models (``rules'') to finally settle on one target ``rule''-- one that might closely mimic a human decision-making task, for example. Probabilistic computing methods like Gen \cite{cite:Gen} and PyMC \cite{cite:PyMC} change a program over time according to a secondary, meta-program that is implicitly or explicitly consistent with Bayesian updating. But in both of these cases, the rules are being updated by a secondary, static program that is set externally. In the case of machine learning, the secondary program is the training methods, in combination with the data being used to train. In the case of probabilistic computing, the secondary program is the Bayesian update method.

But for open-ended systems like living systems, it is unclear \textit{how} rules should change for it to meet the standard of each type of open-endedness. Can we characterize secondary meta-programs such that they relate to the potential kind of open-endedness they could contribute to? In the case of machine learning, the secondary method (training a machine learning model) evolves as well, but at a slower time scale. As the community of machine learning researchers discovers new, efficient methods to train particular model architectures for specific tasks, they introduce new methods for training models such that a certain rule can be discovered faster. Is it necessary to have a rule that changes a rule that changes a rule (three layers of rules) to achieve stronger forms of open-endedness? At present, the answers to these problems are unclear.

\subsection{Imposed constraints}

In the quest to understand how rules for open-ended and living systems change over time, some have explored the relationship between the physical constraints imposed on a system and the dynamics that are allowed under those constraints \cite{cite:Taylor_2021, cite:Montevil_Mossio}. Note the how a constraint is defined in \cite{cite:Montevil_Mossio}:

\begin{quote}
    What do we mean by constraints? In contrast to fundamental physical equations and their underlying symmetries, constraints are contingent causes, exerted by specific structures or dynamics, which reduce the degrees of freedom of the system on which they act. As additional causes, they simplify (or change) the description of the system, and enable an adequate explanation of its behaviour to be provided, an explanation which might otherwise be under-determined or wrongly determined. In describing physical and chemical systems, constraints are usually introduced as external determinations (boundary conditions, parameters, restrictions on the configuration space, etc.), which contribute to determining the behaviour and dynamics of a system, although their existence does not depend on the dynamics on which they act.
\end{quote}

Here, we stress the important distinction between constraints applied to the system \textit{externally} and constraints that \textit{appear fixed} as a result of staying within some subset of the entire possibility state space. For example, the oxygen percentage of our atmosphere acts as a constraint for life on Earth, but that constraint is not upheld externally. Instead, the constraint appears from \textit{within} the system of biological evolution. 

\subsection{The state space is a lie}

The full possibility space for any real system is unthinkably large. But to characterize or describe a system, we, as observers, either \textit{make a decision} on how to best describe a system or \textit{evolve} the ability to sense aspects of a system. By doing so, we construct a symbolic state space that characterizes different aspects (including the temporal dynamics) of that system. This involves selecting relevant variables (color, temperature, pressure, etc) to capture important features necessary for our survival. In other words, the physically embodied way that organisms use to interact with a system determines the structure of the internal representation of that system.

For living structures, it seems like there is something special about these coarse-grained representations. Biological evolution doesn't seem to produce embodiments that produce overly fine-grained representations of the system. While including additional variables would increase the resolution of the system representation, the cost of including these variables may not be worth the amount of information gained by including it \cite{cite:Flack, cite:Walker, cite:Walker12}. But still, living systems seem to continually select embodiments that translate aspects of an system into \textit{information} that enable it to survive \cite{cite:Flack, cite:Walker, cite:Walker12}. So, there seems to be a balance between the need to capture enough information to continue survival, but not too much such that it costs more energy than its worth.

This is exactly the kind of process we wish to explore. We observers never see the full state space of possibilities for any system, but as we interact with real systems as observers, we assign labels to some system macrostates. By doing so, observers construct an internal topology that represents how internal coarse-grained representations of a system transition to other internal coarse-grained representations of a system. An accurate sensory mechanism is a physical embodiment that coarse-grains particular aspects of a system into variables as that observer interacts with the system. Some work suggests that this whole process acts like a compression algorithm on description space for other systems \cite{cite:Flack, cite:Walker, cite:Orozco}. What is the relationship between resulting internal variables, the topology of a ``representation'' of a system, and the physical constraints in an observer's embodiment?

In the next section, we discuss different kinds of constraints based on the kind of behavior a system evolves under while these variables are constrained. We use a toy model to demonstrate how physical constraints result in different representation state spaces. We don't specify whether or not these constrained variables are physically a part of a biological object or in the environment. Instead, we focus on how systems evolve under these constraints. The resulting constrained system dynamics have particular topologies that reflect local, emergent rules. However, we don't address \textit{how} these variables are being constrained-- whether it's from within the system in a homeostatic way, or by an external force doing work on the system. We discuss future directions in establishing a framework that characterizes the relationship between system constraints, constrained system dynamics, and internal dynamics that uphold those constraints \textit{from within} the system.

\section{Examples of constraints as local rules}

\begin{figure*}[t]
\centering
\includegraphics[width=0.6\textwidth]{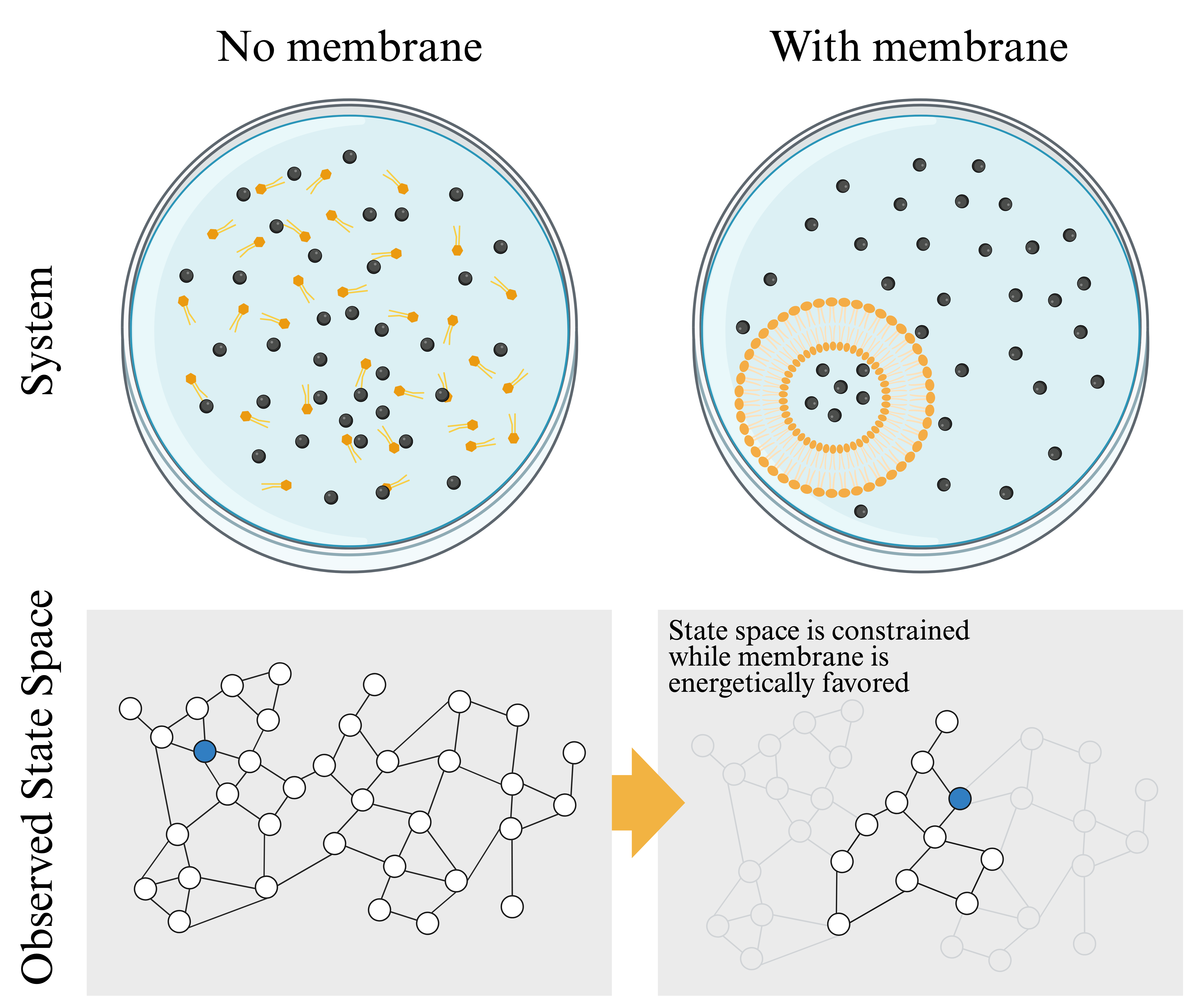}
\caption{An example system of a lipid solution in a petri dish, where a lipid bilayer membrane bubble could form (upper right) or not (upper left). While the membrane bubble is present, it acts as a constraint on the number of possible states the system can physically access. Under some level of description for that system, such as the positions of individual lipids, the dynamics of the real system can be represented as an observed state space with a particular topology, or \textbf{emergent, local rule} (bottom left). The lipids in the solution can be arranged into a lipid bilayer membrane (upper right), either as an energetically favored macrostate or by putting external work onto the system. For either case, doing so constrains the number of possible \textit{observed} states (bottom right) while the constraint (the membrane) is present. These topologies are only used as a toy example and don't reflect the dynamics of an actual, lipid-solution system. Created with BioRender.com.}
\label{fig::membrane}
\end{figure*}

When the laws of physics lead to a system remaining in a smaller region of the much larger possibility space, some observed variables may appear constant (depending on how the variables represent the original system). At least for scientists, the goal of research is to study a system well enough to present its properties and dynamics with certain labels and variables. It often seems like a good set of labels results in a representation of the system's dynamics that \textit{appears} deterministic under those particular labels. This way, we can write the resulting phase space topology\footnote{In physics, a phase space is a theoretical tool used to describe all possible states a system can find itself in, along with all the possible transitions between unique states. Phase spaces are easily visualized in 2 or 3 dimensions, where each axis spans the range of possible variable values to represent the dynamics of that system.} as a system of symbolic equations. Even if a system is proven difficult to approximate symbolically, recent advances in machine learning make it possible to approximate movement along that topology with a trained model. In either case, the resulting representation is a direct reflection of being in the smaller region of the large possibility space. \textbf{The topology (number of states and connections between them) of these smaller regions are local, emergent rules.} If the system were to explore states in another smaller region, the resulting representation of observed system dynamics would be different.

In this section, we discuss constraints (variables that don't change over time) as properties of these smaller regions of a system's possibility state space. While there are other ways of characterizing these smaller spaces (these apparent rules), we use constraints as a simple way to point to different regions of the whole possibility state space. We use the word ``constraint'' to describe \textit{variables that remain fixed} in some region of the state space. However, we note this may be misleading in the usual sense of the word. \textit{This is because it does not necessarily imply whether or not that variable is being held by external forces or generated and maintained by internal ones}. We also justify our interest in calling ``variables that remain fixed in some region of the whole state space of possibilities'' constraints to make an explicit connection to the various physical embodiments that biological objects take to map aspects of their world into internal states.

\begin{figure*}[t]
\centering
\includegraphics[width=0.8\textwidth]{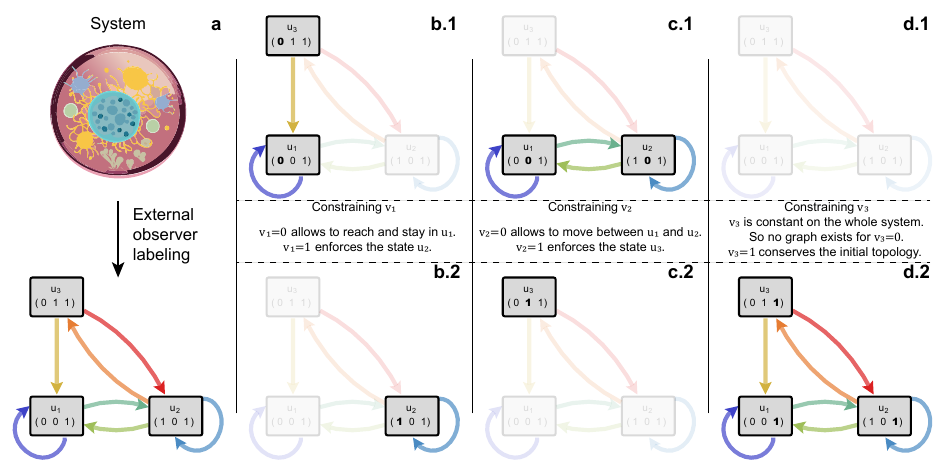}
\caption{Topological alteration of a state graph. a) A system being coarse-grained and receiving arbitrary labels from an external observer. The example macrostate ($u$) labels are vectors of three binary-valued microstate variables denoted $(v_1,v_2,v_3)$. Panels b to d display the topologies of the physically-reachable state graph after constraining a variable of the labeling vector. The nodes and edges that became inaccessible as a consequence of applying a conservation law are displayed with lighter colors. The top row corresponds to setting value 0, while the bottom row corresponds to setting the variable to 1. b) Constraining variable $v_1$. On the one hand (b.1), a conservation law for $v_1=0$ leads the system to state $u_1$. On the other hand (b.2), with $v_1=1$, the system cannot escape from $u_2$. c) Constraining variable $v_1$. In the case $v_1=1$ in c.2), the graph is restricted to $u_3$ and has no possible transitions. The node does exist under a conservation law setting $v_1$ to $1$ but there is no way to reach this node due to the topology of the unconstrained graph. d) Constraining variable $v_3$. This variable is constant and equal to $1$ in the unconstrained graph in a), so a conservation law setting $v_3=0$ such as in panel d.1 is meaningless for this case.}
\label{fig::assembledSimpleEx}
\end{figure*}

\subsection{Lipid membrane}

For example, a petri dish with a lipid solution may or may not form a lipid bilayer membrane bubble (Fig. \ref{fig::membrane}, top). This system is described as some set of variables with particular values (Fig. \ref{fig::membrane}, bottom). The descriptive state space (Fig. \ref{fig::membrane}, lower left) has particular topological properties that are conceptually related to the physical constraints on the system. When the lipid bilayer membrane is present in the system, some set of descriptive variables results in a particular topology of the \textit{physically accessible} states under that constraint. In the bottom row of Fig. \ref{fig::membrane}, the nodes of this topology are labeled with vectors (see Fig. \ref{fig::assembledSimpleEx}). The membrane constrains the number of system possibilities into a small subset of that state space (Fig. \ref{fig::membrane}, lower right). The constraint of the membrane not only imposes a physical constraint on the movement of the lipids, but it also represents a particular subset of an observer's representation of that system as a lipid-position possibility state space. As long as the membrane remains intact, the \textit{physically realized} states of the system remain constrained. 

In this example, three things are happening: (1) A set of variables describes the dynamics of the system into some ``observed'' state space with some particular topology. The states are each labeled by a set of variables that represent aspects and properties of individual lipid coordinates. (2) When the membrane constraint is present in the system, the number of physically accessible states shrinks into a smaller subset. For some observed state spaces, the nodes in the smaller subset might have one or more variables with values that are constant for these particular states. For the bubble, these variables might represent the number of adjacent lipids to any single lipid, which remains fairly constant while they are arranged in a lipid bilayer membrane. (3) The membrane constraint is only present as a result of the laws of physics of the system (being energetically favorable), or by putting external work into the system (physically holding the lipids together through some mechanism). Note how for this example, the constraint being present is not a result of the internal dynamics of the bubble. 

\subsection{Membrane of a living cell}

In more complex, living cells, whatever is contained in the cell also maintains the very membrane that constrains the contents of the cell. The loss of the ``living state'' for the cell results in the mechanisms that maintain the membrane to cease, thus dissolving the membrane. So for living objects like cells, it seems that constraints that define the object are also present via mechanisms that are only dynamically possible while that constraint is present. In other words, the cell membrane \textit{conditionally constrains} mechanisms that allow it to persist \cite{cite:Varela, cite:Maturana}.

The stability of a cell membrane acts like a conditional conservation law because the membrane traps objects inside the cell. While the cell membrane itself is a constraint on the positions of lipids, the contents contained inside the lipid cell are constrained \textit{conditionally} as a result of the membrane. In other words, the membrane constraint suddenly implies a second, conditionally constrained set of variables in a descriptive state space that describes the location of the trapped objects. In addition, the state of the membrane constraint is maintained by the dynamics of objects it has trapped. For this example, this constraint is a direct result of dynamics from \textit{within the system}, which also \textit{conditionally constrains} the contents of the cell.

\subsection{Autocatalytic sets}

Autocatalytic sets are an example of chemical systems with self-creating dynamics. The conditions needed to perform particular chemical dynamics are produced as a result of those dynamics. Like the membrane of a living cell, the physical conditions needed for autocatalysis are maintained and upheld as a result of the dynamics of the constrained state space.

Often considered a primitive, pre-biotic form of autopoiesis, researchers have studied autocatalytic reaction networks characterized by chemical reaction dynamics that display a kind of biological functionality \cite{cite:Baum, cite:Peng}. Here, the autocatalytic functionality is a reflection of the topology of the chemical system evolving under particular constraints.

\subsection{Autocatalytic networks}

A reaction network is autocatalytic in a set of species if it consumes the set while also producing it at an even higher rate \cite{cite:Blokhuis}. Here, the conditions needed to perform a particular chemical dynamic are produced as a result of those dynamics. Like the membrane of a living cell, the physical conditions needed for autocatalysis are maintained and upheld as a result of the dynamics of the constrained state space. In dynamics where the autocatalytic species are no longer spontaneously produced by the environment, an extinction event of a particular autocatalytic set permanently constrains the system to a smaller subspace. On the other hand, if the \textit{food (waste)} needed (produced) by the autocatalysts \cite{cite:Gagrani} is provided (removed) at a sufficiently large rate, the same system will be constrained to an attractor basin or a growth regime very far from equilibrium.

Often considered a primitive, pre-biotic form of autopoiesis, researchers have studied autocatalytic reaction networks characterized by chemical reaction dynamics that display a kind of biological functionality \cite{cite:Baum, cite:Peng_2}. It is shown that ecosystems of autocatalytic subnetworks can recover familiar ecological dynamics. Moreover, it is argued that a population of interacting ecosystems (compartmentalized or diffusing on a lattice) can show selection toward the discovery of autocatalytic subnetworks that increase the growth rate of the ecosystem they constitute. At both levels, the autocatalytic functionality is a reflection of the topology of the chemical system evolving under particular constraints due to its own dynamics.

\subsection{Feedback-based canalization}

The way that cracks form and grow in solids is a non-biological example of how the creation of constraints can create a feedback mechanism that reinforces the growth of that crack. Given a solid, there are lots of potential cracks that could exist, but once any initial crack forms, it ends up concentrating all of the available stress on its tip. Due to this stress, that particular crack continues to form, rather than just other cracks also randomly forming. Because of the physical properties of an initial crack, the system is energetically favored to reinforce that one particular crack, since crack propagation is fundamentally a non-equilibrium process.

\subsection{Living systems}

But unlike membranes of bubbles and autocatalytic sets, living systems seem to maintain these constraints as a result of the internal system \textit{doing work} to physically maintain the constraints that define them (homeostatic processes, for example). These kinds of biological ``endo-physics'' don't physically realize the other possible states in the subspace because they actively spend energy to \textit{avoid} them. 

Yet, when we describe biological systems, we often refer to their ability to constantly explore new, innovative states. But we suggest describing living objects according to their ability to avoid certain states, as per the constraints that define their physical embodiment. As with the case of autocatalytic sets and membranes of living cells, the topology of the \textit{constrained} state space results in functionality. Given that perspective, we seek to identify the relationship between different types of constraints and resulting functionality that may be related to well-known functions of living systems \cite{cite:Wong_etal}. In the next section, we use constraints (fixed variables) on graphs (nodes connected by directed edges) to demonstrate how the resulting, constrained possibility state space depends on (1) the value of the fixed variable and (2) an observed system's initial state space, especially the transitions between states (global laws).

\section{The effects of constraints on a system's possibility state space}

Here, we see what happens to the physically-accessible state space for a system that is somehow constrained. For these examples, we \textit{don't} specify whether these variables are fixed due to internal system dynamics or due to external forces acting on the system. We also don't specify if a fixed variable is a part of a biological object or is a part of the object's environment. Instead, we never define an object in any of these systems.

First, we start with a different example dynamical system than the lipid bilayer membrane from Fig. \ref{fig::membrane}. Given some description of a new dynamical system, shown in Fig. \ref{fig::assembledSimpleEx}.a, we can systematically explore the effects of constraining all three microstate variables, regardless of how those variables are physically set to remain in those values. By being able to see how the physically accessible possibility state space (Fig. \ref{fig::assembledSimpleEx}, top and bottom rows of panels b to d) changes as a result of that constraint, we conceptually connect topological properties with \textit{apparent} functionality. These functions, which we call local, emergent rules, become physically realized when particular variables are constrained. While we don't yet know how to make these connections, we discuss various aspects of this approach more broadly.

Namely, we can already identify patterns in this simple example. Functions that could emerge from being in a sole state are particularly represented in Fig. \ref{fig::assembledSimpleEx} when constraining variable $v_1$ to either $0$ or $1$ (Fig. \ref{fig::assembledSimpleEx}.b.1 and Fig. \ref{fig::assembledSimpleEx}.b.2, respectively), or variable $v_2$ to $1$. Note that in the case $v_1=0$, the state $u_1$ becomes the final state after transiting through $u_3$ first. We can imagine extending this structure to a longer state chain in which case the relevancy of the emerging function for the system would lie in the specific sequence of state stabilized by the constraint. In the case of $v_1=1$, however, the function would be a one-to-one map between a constraint and a unique state maybe suggesting the criticality of this state for the biological system. Indeed, the system cannot waste energy roaming the state space in such cases and must remain in place. The case of $v_2=1$ is yet another possibility where the state does exist but no dynamic can be observed related to it. This might suggest that the state is reachable in another graph (system description) that would share variable $v_2$ but not $v_1$ and $v_3$. On the contrary to the case $v_1=0$ (Fig. \ref{fig::assembledSimpleEx}.b.1), the case $v_2=0$ yields a constrained graph with a loop between two states. This type of structure might indicate a lower criticality of the function of the biological system or the need for a dual solution to the same problem. Finally, there is the case of constant variables such as $v_3$ in the observable graph (see in Fig. \ref{fig::assembledSimpleEx}.a that the third component of the vector is always equal to 1). In such cases, constraints on the variable have either the effect to turn off the whole observed state space, or to turn it on, unaltered.

We propose three different classes of constraints based on how quickly or readily they can change over time (or more generally, over a walk along the state space). Some variables, like Planck's constant, are \textbf{completely conserved}. These variables are a direct property of the laws of physics that determine the paths between arbitrarily labeled states.  More subtly, some variables define the local dynamics of a system such that the time derivative of other variables is zero. While these variables themselves are not conserved, they \textbf{conditionally conserve} other variables in the system. To these two types, we add the possibility to be \textbf{quasi-conserved}. This quality indicates that the constraint does not perfectly implement a subset of states where it is satisfied but, at least, it is only possible to escape from the subset that perfectly respects the constraint very slowly over time relative to other variables. This includes co-evolutive phenomena like nitrogen fixation in the soil or the amount of oxygen in the atmosphere.

\section{Future directions and discussion} 

In this discussion paper, we have examined the relationship between constraints and functionality over a system's state space represented as a directed graph. We argue that the resulting, accessible state space under a constraint is akin to a local, emergent conservation law or apparent ``rule'' that some part of the system evolves under. While we have identified different effects of constrained variables (ones that produce rules that are completely conserved, quasi-conserved, and conditionally conserved), there is still much more that needs to be explored to translate constrained topologies into functions important for living systems-- particularly how emergent ``rules'' contribute to open-ended system dynamics. Which types of constrained variables are most likely to lead to open-ended system dynamics? Since these topologies also depend on the unconstrained representation of a system, we also want to understand the relationship between constrained topologies and non-constrained ones. For example, how do system topologies differ between isolated systems and systems where external work is being done on them? Furthermore, is there an analytical approach to measuring possible constraints from within the system?

Understanding how biological systems and other open-ended systems continually generate new, innovative rules over time has deep implications for many fields and problems, all of which center on understanding open-endedness. These include Artificial Life, characterizing diverse intelligences, understanding biological processes and major evolutionary transitions, designing machine learning architectures capable of self-teaching, and more. We hope that by deepening our understanding of how physical constraints are added, changed, and removed via the global laws of physics leads to emergent, local rules, we gain a small step closer to understanding mechanisms that drive open-endedness. We hope this discussion serves as a springboard to explore approaches in this direction, as has been previously highlighted by \cite{cite:Taylor_2021, cite:Montevil_Mossio}.

\section*{Acknowledgments}

We would like to thank Nicholas Guttenberg for several detailed discussions and brainstorming that led to this paper. We would also like to thank Martin Biehl, Joseph Austerweil, Tim Taylor, Hector Zenil, Kenneth Stanley, Lisa Soros, and Lana Sinapayen for their feedback, guidance, and discussions for many of the ideas discussed throughout the paper. PG was supported by the National Science Foundation, Division of Environmental Biology (Grant No: DEB-2218817). Finally, we thank the reviewers for their helpful comments, feedback, and suggestions.







\begin{thebibliography}{00}

\bibitem{cite:Taylor_2021} T. Taylor, ``Evolutionary Innovation Viewed as Novel Physical Phenomena and Hierarchical Systems Building'', presented at the \textit{Fourth Workshop on Open-Ended Evolution} (OEE) at ALIFE 2021, Prague/Online, July 2021, https://doi.org/10.48550/arXiv.2107.09669

\bibitem{cite:Bartlett_Wong} S. Bartlett and M.L. Wong, ``Defining Lyfe in the Universe: From Three Privileged Functions to Four Pillars'', Life, vol. 10, pp 42, April 2020

\bibitem{cite:Wong_Bartlett_Chen_Tierney} M.L. Wong,  S. Bartlett, S. Chen, and L. Tierney ``Searching for Life, Mindful of Lyfe’s Possibilities'', Life, vol. 12, pp 783, June 2022

\bibitem{cite:Wong_etal} M. L. Wong, C. E. Cleland, D. Arend Jr., S. Bartlett, H. J. Cleaves II, H. Demarest, et al., ``On the roles of function and selection in evolving systems'', PNAS, vol. 120(43), pp e2310223120, October 2023

\bibitem{cite:Sharma} A. Sharma, D. Czégel, M. Lachmann, C. P. Kempes, S. I. Walker, and L. Cronin, ``Assembly theory explains and quantifies selection and evolution'', Nature, vol. 622, pp 321--328,  Oct. 2023

\bibitem{cite:Kim} H. Kim, G. Valentini, J. Hanson, and S. I. Walker, ``Informational architecture across non-living and living collectives'', Theory Biosci., vol. 140, pp 325--341, Feb. 2021

\bibitem{cite:Kempes} C. P. Kempes and D. C. Krakauer, ``The Multiple Paths to Multiple Life'', J. Mol. Evol., vol. 89(7), pp. 415–426, Aug. 2021

\bibitem{cite:Dupre} J. Dupré, ``The Metaphysics of Biology'', Elements in the Philosophy of Biology, May 2021

\bibitem{cite:Witzany} G. Witzany, ``What is Life?'', Front. Astron. Space Sci., vol. 7, March 2020

\bibitem{cite:Krakauer} D. Krakauer, N. Bertschinger, E. Olbrich, J. C. Flack, and N. Ay, ``The information theory of individuality'' Theory Biosci., vol. 139(2), pp 209-–223, Jun. 2020

\bibitem{cite:Jeancolas} C. Jeancolas, C. Malaterre, and P. Nghe, ``Thresholds in Origin of Life Scenarios'', iScience, vol. 23(11), pp 101756, November 2020

\bibitem{cite:Pattee} H. H. Pattee and H. Sayama, ``Evolved Open-Endedness, Not Open-Ended Evolution'' Artif. Life, vol. 25(1), pp 4-–8, April 2019

\bibitem{cite:Packard} N. Packard et al., ``An Overview of Open-Ended Evolution: Editorial Introduction to the Open-Ended Evolution II Special Issue'', Artif. Life, vol. 25(2), pp 93-–103, May 2019

\bibitem{cite:Flack} J. Flack, ``Life’s Information Hierarchy'', in \textit{From Matter to Life: Information and Causality}, G. F. R. Ellis, P. C. W. Davies, and S. I. Walker, Eds., Cambridge: Cambridge University Press, March 2017, pp 283-–302. 


\bibitem{cite:stanley} K. O. Stanley, ``Why Open-Endedness Matters'', Artif. Life, 25(3), pp 232-–235, August 2019

\bibitem{cite:Song} A. Song, ``A little taxonomy of open-endedness'', presented at the \textit{Workshop on Agent Learning in Open-Endedness} (ALOE) at ICLR 2022, April 2022.

\bibitem{cite:Packard19} N. Packard, Mark A Bedau, A. Channon, T. Ikegami, S. Rasmussen, K. Stanley, and T. Taylor, ``Open-Ended Evolution and Open-Endedness: Editorial Introduction to the Open-Ended Evolution I Special Issue'', Artif. Life, vol. 25(1), pp 1-–3, 2019

\bibitem{cite:Bedau} M. A. Bedau, N. Gigliotti, T. Janssen, A. Kosik, A. Nambiar, and N. Packard, ``Open-Ended Technological Innovation'', Artif. Life, vol. 25(1), pp 33-–49, April 2019

\bibitem{cite:stanley17} K. Stanley, J. Lehman, and L. Soros, ``Open-endedness: The last grand challenge you’ve never heard of.'' Available at https://www.oreilly.com/radar/open-endedness-the-last-grand-challenge-youve-never-heard-of/ (2017/12/19)

\bibitem{cite:Taylor_2016} T. Taylor,  M. Bedau, A. Channon, D. Ackley, W. Banzhaf, G. Beslon, et al., ``Open-Ended Evolution: Perspectives from the OEE1 Workshop in York'', Artif. Life, vol. 22(3), pp 408-–423, August 2016.

\bibitem{cite:Soros} L. Soros and K. O. Stanley, ``Identifying necessary conditions for open-ended evolution through the artificial life world of Chromaria'', in \textit{Proceedings of the 2014 Artificial Life Conference} (ALIFE 2014), pp. 793–800, July 2014

\bibitem{cite:Bartlett_Louapre} S. Bartlett and D. Louapre, ``Provenance of life: Chemical autonomous agents surviving through associative learning'', Phys. Rev. E 106, 034401, September 2022

\bibitem{cite:Montevil_Mossio} M. Montévil and M. Mossio, ``Biological organisation as closure of constraints'', J Theor Biol, vol. 372, pp. 179--191, May 2015

\bibitem{cite:Taylor_2019} T. Taylor, ``Evolutionary Innovations and Where to Find Them: Routes to Open-Ended Evolution in Natural and Artificial Systems'', Artif. Life, vol.25(4), pp 207--224, May 2019

\bibitem{cite:banzhaf} W. Banzhaf, B. Baumgaertner, G. Beslon, R. Doursat, J. A. Foster, B. McMullin, et al., ``Defining and simulating open-ended novelty: requirements, guidelines, and challenges'', Theory in Biosciences, vol. 135, pp 131–-161, May 2016

\bibitem{cite:Hintze} Arend Hintze, ``Open-Endedness for the Sake of Open-Endedness'' Artif. Life, 25(2), pp 198-–206, May 2019

\bibitem{cite:Wang} R. Wang, J. Lehman, J. Clune, and K. O. Stanley, ``Paired Open-Ended Trailblazer (POET): Endlessly Generating Increasingly Complex and Diverse Learning Environments and Their Solutions.'', arXiv, February 2019, Available at https://arxiv.org/abs/1901.01753

\bibitem{cite:Wang20} R. Wang, J. Lehman, A. Rawal, J. Zhi, Y. Li, J. Clune, and K. O. Stanley, ``Enhanced POET: Open-ended Reinforcement Learning through Unbounded Invention of Learning Challenges and their Solutions'', in \textit{Proceedings of the 37th International Conference on Machine Learning, PMLR}, pp 9940-–9951, November 2020 

\bibitem{cite:Dolson} E. L. Dolson, A. E. Vostinar, M. J. Wiser, and C. Ofria, ``The MODES Toolbox: Measurements of Open-Ended Dynamics in Evolving Systems'', Artif. Life, 25(1), pp 50-–73, 2019

\bibitem{cite:Horibe} K. Horibe, K. Suzuki, T. Horii, H. Ishiguro, ``Exploring the Adaptive Behaviors of Particle Lenia: A Perturbation-Response Analysis for Computational Agency.'', in \textit{Proceedings of the 2023 Artificial Life Conference} (ALIFE 2023), pp 40, July 2023.

\bibitem{cite:Ackley} D. Ackley, ``Indefinite scalability for living computation'', in \textit{Proceedings of the 13th AAAI Conference on Artificial Intelligence} (AAAI-16), vol. 30(1), February 2016.

\bibitem{cite:Ackley14} D. Ackley and T. Small, ``Indefinitely scalable computing= artificial life engineering'', in \textit{Proceedings of the 2014 Artificial Life Conference} (ALIFE 2014), pp. 606–613, July 2014

\bibitem{cite:PyMC} O. Abril-Pla, V. Andreani, C. Carroll, L. Dong, C. J. Fonnesbeck, M. Kochurov, et al., ``PyMC: A Modern and Comprehensive Probabilistic Programming Framework in Python'', PeerJ Comput. Sci., vol. 9, pp e1516, September 2023

\bibitem{cite:Gen} M. F. Cusumano-Towner, F. A. Saad, A. K. Lew, and V. K. Mansinghka. ``Gen: a general-purpose probabilistic programming system with programmable inference'', in \textit{Proceedings of the 40th ACM SIGPLAN Conference on Programming Language Design and Implementation} (PLDI 2019), pp 221–236, June 2019

\bibitem{cite:Adams} A. Adams, H. Zenil, P. C. W. Davies, and S. I. Walker, ``Formal Definitions of Unbounded Evolution and Innovation Reveal Universal Mechanisms for Open-Ended Evolution in Dynamical Systems'', Sci. Rep., vol. 7, 2017

\bibitem{cite:Liard} V. Liard, D. P. Parsons, J. Rouzaud-Cornabas, and G. Beslon, ``The Complexity Ratchet: Stronger than Selection, Stronger than Evolvability, Weaker than Robustness'', Artif. Life, vol. 26(1), pp 38-–57, April 2020

\bibitem{cite:Tennie} C. Tennie, J. Call, and M. Tomasello, ``Ratcheting up the ratchet: on the evolution of cumulative culture''. Philos. Trans. R. Soc. Lond. B. Biol. Sci., vol. 364(1528), pp 2405--2415. August 2009


\bibitem{cite:Baltieri} M. Baltieri, H. Iizuka, O. Witkowski, L. Sinapayen, and K. Suzuki. ``Hybrid Life: Integrating biological, artificial, and cognitive systems'', WIREs Cognitive Science, vol. 14(6), pp e1662. July 2023

\bibitem{cite:Smith} E. Smith, B. S. Harrison, and J. L. Andersen. ``Rules, hypergraphs, and probabilities: the three-level analysis of chemical reaction systems and other stochastic stoichiometric population processes.'', bioRxiv, December 2023, Available at https://www.biorxiv.org/content/10.1101/2023.12.11.571120v1.full

\bibitem{cite:Bedau97} M. A. Bedau, ``Weak Emergence'', Noûs, vol. 31, pp. 375–399, 1997

\bibitem{cite:Bedauetal97} M. A. Bedau, E. Snyder, C. T. Brown, N. H. Packard, and others, ``A comparison of evolutionary activity in artificial evolving systems and in the biosphere''; in \textit{Proceedings of the 4th European Conference on Artificial life} (ECAL97), pp. 125–-134, July 1997

\bibitem{cite:Lavin} A. Lavin, D. Krakauer, H. Zenil, J. Gottschlich, T. Mattson, J. Brehmer, et al., ``Simulation Intelligence: Towards a New Generation of Scientific Methods,” arXiv, November 2022, Available at http://arxiv.org/abs/2112.03235

\bibitem{cite:Orozco} S. Hernández-Orozco, F. Hernández-Quiroz, and H. Zenil, ``Undecidability and Irreducibility Conditions for Open-Ended Evolution and Emergence'', Artif. Life, vol. 24(1), pp 56-–70, February 2018

\bibitem{cite:Mitchell} M. Mitchell, ``Why AI is harder than we think'', in \textit{Proceedings of the Genetic and Evolutionary Computation Conference} (GECCO 2021), pp 3, June 2021

\bibitem{cite:Walker} S. I. Walker and P. C. Davies, ``The algorithmic origins of life'', J. R. Soc. Int., vol. 10(79), pp 20120869, February 2013.

\bibitem{cite:Walker12} S. I. Walker, L. Cisneros, and P. C. Davies, ``Evolutionary transitions and top-down causation'', in \textit{Proceedings of the 2012 Artificial Life Conference} (ALIFE 2012), pp. 283–-290, July 2012.

\bibitem{cite:Varela} F. G. Varela, H. R. Maturana, and R. Uribe, ``Autopoiesis: The organization of living systems, its characterization and a model'', Biosystems, vol. 5(4), pp 187-–196, May 1974

\bibitem{cite:Maturana} H. R. Maturana and F. J. Varela, ``Autopoiesis and Cognition: The Realization of the Living'', 2nd ed., ser. Boston Studies in the Philosophy and History of Science, Dordrecht Netherlands: Springer, April 1980, vol. 42


\bibitem{cite:Blokhuis} A. Blokhuis, D. Lacoste, and P. Nghe. ``Universal motifs and the diversity of autocatalytic systems'', PNAS, vol. 117(41), pp 25230-25236, September 2020

\bibitem{cite:Gagrani} P. Gagrani, V. Blanco, E. Smith, and D. Baum. ``The geometry and combinatorics of an autocatalytic ecology in chemical and cluster chemical reaction networks'' arXiv, November 2023, available at https://arxiv.org/abs/2303.14238

\bibitem{cite:Baum} D. A. Baum, Z. Peng, E. Dolson, E. Smith, A. M. Plum, and P. Gagrani, ``The ecology–evolution continuum and the origin of life'', J. R. Soc. Int., vol. 20(208), pp 20230346, November 2023

\bibitem{cite:Peng} Z. Peng, J. Linderoth, and D. A. Baum, ``The hierarchical organization of autocatalytic reaction networks and its relevance to the origin of life'', PLOS Comp. Biology, vol. 18(9), pp e1010498, September 2022.

\bibitem{cite:Peng_2} Z. Peng, A. M. Plum, P. Gagrani, and D. A. Baum, ``An ecological framework for the analysis of prebiotic chemical reaction networks'', J. Theo. Bio., vol. 507, pp 110451, December 2020

\bibitem{cite:Hernandez-Orozco} S. Hernández-Orozco, N. A. Kiani, and H. Zenil, “Algorithmically probable mutations reproduce aspects of evolution, such as convergence rate, genetic memory and modularity,” Royal Society Open Science, vol. 5, no. 8, p. 180399, Aug. 2018, doi: 10.1098/rsos.180399.

\bibitem{cite:Zenil} H. Zenil, N. A. Kiani, M. Shang, and J. Tegnér, “Algorithmic Complexity and Reprogrammability of Chemical Structure Networks,” Parallel Process. Lett., vol. 28, no. 01, p. 1850005, Mar. 2018, doi: 10.1142/S0129626418500056.


\end{thebibliography}
\end{document}